\documentclass[12pt,preprint]{aastex}

\newcommand{\myemail}{guerrero@astro.iag.usp.br}
\newcommand{\sr}{R_{\odot}}
\newcommand{\degree}{^{\circ}}

\shorttitle{eta-quenching in Babcock-Leighton dynamos}
\shortauthors{Guerrero, Dikpati \and Dal Pino}

\begin{document}


\title{THE ROLE OF DIFFUSIVITY QUENCHING IN FLUX-TRANSPORT DYNAMO
MODELS}


\author{Gustavo Guerrero\altaffilmark{1} }
\affil{Astronomy Department, IAG, Universidade de Sao Paulo,
    Brazil}
\email{\myemail}

\author{Mausumi Dikpati}
\affil{High Altitude Observatory, NCAR}
\email{dikpati@ucar.edu}

\and

\author{Elisabete M. de Gouveia Dal Pino}
\affil{Astronomy Department, IAG, Universidade de Sao Paulo,
  Brazil}
\email{dalpino@astro.iag.usp.br}


\altaffiltext{1}{The major part of this work was done during Gustavo
Guerrero's visiting appointment at High Altitude Observatory, NCAR
in summer of 2008.}


\begin{abstract}

In the non-linear phase of a dynamo process, the back-reaction of 
the magnetic field upon the turbulent motion results in a decrease 
of the turbulence level and therefore in a suppression of both
the magnetic field amplification (the $\alpha$-quenching effect) and 
the turbulent magnetic diffusivity (the $\eta$-quenching effect). While 
the former has been widely explored, the effects of $\eta$-quenching 
in the magnetic field evolution have rarely been considered. In this 
work we investigate the role of the suppression of diffusivity in a 
flux-transport solar dynamo model that also includes a non-linear 
$\alpha$ quenching term. Our results indicate that, although for 
$\alpha$-quenching the dependence of the magnetic field amplification 
with the quenching factor is nearly linear, the magnetic field response 
to $\eta$-quenching is non-linear and spatially non-uniform. We 
have found that the magnetic field can be locally amplified in this 
case, forming long-lived structures whose maximum amplitude can be 
up to $\sim2.5$ times larger at the tachocline and up to $\sim2$ 
times larger at the  center of the convection zone than in models 
without quenching. However, this amplification leads to 
unobservable effects and to a worse  distribution of the magnetic 
field in the butterfly diagram. Since the dynamo cycle period 
increases when the efficiency of the quenching increases, we have 
also explored whether the $\eta$-quenching can cause a 
diffusion-dominated model to drift into an advection-dominated 
regime. We have found that models undergoing a large suppression 
in $\eta$ produce a strong segregation of magnetic fields that may 
lead to unsteady dynamo-oscillations. On the other hand, an initially 
diffusion-dominated model undergoing a small suppression in $\eta$ 
remains in the diffusion-dominated regime.

\end{abstract}


\keywords{MHD---sun: magnetic fields}

\section{INTRODUCTION}

Over the past half a century, since the development of the first solar
dynamo model by Parker (1955), significant investigations have been
performed to find the saturation mechanism that would limit the growth
of a dynamo. Such mechanisms are likely to include feedback processes, 
such as the back-reaction of magnetic fields on the flow fields,
including mean flows (differential rotation and meridional
circulation), as well as turbulent flows.

The feedback process that has been most extensively studied in the
mean-field electrodynamics is the back-reaction of magnetic fields
on the helical part of the turbulent flow. 
This process, often known as $\alpha$-quenching,
was first believed to be mainly due to the back-reaction of the
magnetic field on the convection, causing the suppression of kinetic
helicity, and hence quenching the inducing effects of the turbulent
electromotive force \citet{stix72}. Currently it is believed that
the saturation process in the non-kinematic regime occurs due to
the reduction of the kinetic $\alpha$-effect caused by a magnetic
contribution of opposite sign, coming from the equation of the 
$\alpha$. This contribution appears asa product of the magnetic
helicity conservation constraint. There exists
a large literature on $\alpha$-quenching and, instead of detailed
review, we refer to the following papers on this topic: \citet{kraichnan79,
catt91, gd94, by95, ch96, bd97, fbc99, bf01, fb02, bs05}.

However, the back-reaction due to induced magnetic field on the
mirror-symmetric non-helical part of the turbulent flow is a relatively
less explored subject. Such back reaction has the effect of reducing
the eddy diffusivity. This back-reaction process has been named
the $\eta$-quenching, first derived by \citet{rs75} using mean-field
electrodynamics.

Later, the role of $\eta$-quenching in a dynamo has been investigated
by several authors \citep{rkks94, tobias96}, primarily in the
context of $\alpha\Omega$-type stellar dynamo models. \citet{rkks94}
incorporated $\eta$-quenching in one-dimensional $\alpha\Omega$ type
and two-dimensional $\alpha^2 \Omega$ type stellar dynamo models.
For supercritical dynamo regimes, \citet{rkks94} found that in
one-dimensional dynamo models with $\eta$-quenching, the field
strength increased a bit, but not much, whereas the cycle period
decreased significantly. By contrast, for their
two-dimensional stellar dynamo models, \citet{rkks94} found a
significant field amplification, almost two times more magnetic
field was produced with $\eta$-quenching, but the cycle period
did not change much from that without $\eta$-quenching. While 
field amplification due to $\eta$-quenching is intuitively expected,
the change in dynamo cycle period, $T$, will depend on how $T$ is
determined in different classes of dynamo models. For example, in
the convection zone $\alpha\Omega$ dynamo models, the cycle frequency
($\omega$) follows $\omega \propto \Omega^{0.5} \eta^{0.5}$; hence an
expected increase in cycle period with $\eta$-quenching.

By employing an $\alpha\Omega$ type interface dynamo model,
\citet{tobias96} found that for weak magnetic fields, the diffusivity
is not much quenched, and the dynamo solutions are not influenced
by the presence of $\eta$-quenching. In the case of strong magnetic
fields, \citet{tobias96} showed that the diffusivity near the base of
the convection zone can be so heavily quenched that the fields can be
trapped there without making their buoyant escape towards the solar
surface.

In a more recent calculation \citet{gr05} showed, by solving the
induction equation for the toroidal magnetic field component
including the back-reactions of magnetic fields on the turbulent
diffusivity as well as on the shear, that there exists a competition
between a field amplification by $\eta$-quenching and a field
reduction due to the fact that the the Lorentz force feedback on
the shear moves the latitudinal shear layer away from the
mid-latitudes. However, \citet{gr05} argued that a significant
field amplification might be possible if the latitudinal shear
is replenished in a much shorter time-scale compared to the solar
cycle.

There is an effect common to all the above studies which either
solved an $\alpha\Omega$ convection zone dynamo \citep{rkks94},
an interface dynamo \citep{tobias96}, or an induction equation for
the toroidal magnetic field component \citep{gr05} -- there is field
amplification due to $\eta$-quenching. But the influence of $\eta$-quenching
on solar cycle features, namely the butterfly diagram and the evolutionary
pattern of magnetic fields in the convection zone and tachocline,
has not yet been explored.

Our aim here is to simulate a Babcock-Leighton flux-transport dynamo
including the back-reactions of magnetic fields on both the helical
and non-helical parts of the turbulent flow, i.e.  by including both
the $\alpha$-quenching and $\eta$-quenching. We specifically seek the
answers to the following questions: (i) what is a characteristic value
of field amplification due to $\eta$-quenching in a Babcock-Leighton
flux-transport dynamo model? (ii) How is the butterfly diagram changed
or modified due to $\eta$-quenching? (iii) Where in the solution domain
that extends from pole-to-equator in latitude and from the tachocline to 
the solar surface in radial extent does $\eta$-quenching have the most
effect? (iv) Is the conveyor-belt mechanism preserved,
or does it break down if $\eta$ is more and more quenched? (v) To
what extent can the $\eta$-quenching affect the dynamo cycle period
which is primarily determined by the meridional circulation in this
class of models? (vi) Since the $\alpha$-quenching saturates the
growth of the dynamo field and the $\eta$-quenching works in
amplifying the field, how does the dynamo behave in presence of the
competition between these two quenching mechanisms? (vii) Can the
$\eta$-quenching take a diffusion-dominated dynamo into the 
advection-dominated regime?

In the next section we present the formulation of the model, including
the prescriptions of the dynamo ingredients, such as the
velocity fields (differential rotation and meridional circulation),
diffusivity profile with $\eta$-quenching formula, and the
$\alpha$-effect profile. We present the solution method, boundary
conditions and initial conditions in section 3 and the detailed
description of our results in section 4. We conclude in section 5.

\section{MATHEMATICAL FORMULATION}

In the mean field approximation, the MHD induction equation that
governs the evolution of the large scale magnetic field is:

\begin{equation}\label{eq1}
\frac{\partial {\bf B}}{\partial t} = \nabla \times [{\bf U} \times
    {\bf B} + {\bf \cal{E}} - \eta_T \nabla \times {\bf B} ],
\end{equation}
where ${\bf B}$ is the mean magnetic field and ${\bf U}$ is the large
scale velocity field, $\eta_T$ is the magnetic diffusivity and
{\bf $\cal{E}$} is the electromotive force, ${\bf u} \times {\bf b}$,
that represents the contribution of the small scale fluctuations upon
the large scales. There have been several previous works in the
literature where the effects of the latter term have been studied in
detail \citep{kukeretal01,bonnetal02,ketal06b,gdp08}. These 
studies include the contribution in the mean-field dynamo equation 
explicitly from different parts of magnetic diffusivity tensor arising 
from small-scale turbulence along and perpendicular to the rotation axis.
In the present work since we are concerned in exploring the effects of the 
$\eta$-quenching, we are going to neglect these detailed effects 
of {\bf $\cal{E}$} other than the $\alpha$-effect and the isotropic
turbulent diffusivity.

Working in spherical coordinates and assuming spherical
symmetry, we can write ${\bf B_p}$$=$$\nabla \times (A\hat{e}_{\phi})$
and $B \hat{e}_{\phi}$, as the poloidal and toroidal components of the
magnetic field, respectively, and considering ${\bf u_p}$ and ${\bf
  \Omega} r \sin \theta$ as 
the meridional velocity field and the differential rotation,
respectively, then we can split eq. (\ref{eq1}) in the
following pair of coupled partial differential equations for
$A$ and $B$:

\begin{eqnarray}
\frac{\partial A}{\partial t}+\frac{1}{s}[{\bf u_P}
\cdot\nabla](sA)=\eta_T\left(\nabla^2-\frac{1}{s^2}\right)A +
S_1(r,\theta,B)\label{eq2} \quad ,\\
\frac{\partial B}{\partial t}+\frac{1}{r}\left[\frac{\partial}{\partial
    r}(r u_r B)+\frac{\partial}{\partial
    \theta}(u_{\theta} B)\right]\label{eq3}=s({\bf
  B_p} \cdot \nabla) \Omega\\\nonumber
-\left[\nabla \eta_T \times (\nabla \times
B \hat{e}_{\phi})\right]_{\phi}
+\eta_T \left(\nabla^2-\frac{1}{s^2}\right)B\nonumber \quad,
\end{eqnarray}

\noindent where $s=r sin\theta$. The other terms will be
described in detail in the next paragraphs.

\subsection{The velocity field}

The velocity field is one of the most important ingredients in
the kinematic solar dynamo. In the axisymmetric regime of a
Babcock-Leighton flux-transport dynamo, the
velocity field can be split into two large-scale mean-flow
components in the azimuthal ($\phi$) and the meridional ($r,\theta$)
directions. The azimuthal component is the differential rotation,
which is the responsible for the generation of the toroidal
fields. The ($r,\theta$) component is the meridional flow
which transports the magnetic flux first poleward at the surface
and then equatorward at the bottom of the convection zone, as in a
conveyor belt. This ingredient also plays a crucial role in
determining the dynamo cycle period \citep{wang91, dc99, kukeretal01}
and the memory of the Sun's past magnetic fields \citep{ddg06}
if the advection dominates over the diffusion in determining the
characteristic time-scale of the system.

Recent observational developments gave us the access to an accurate
and detailed profile of the differential rotation in the entire
convection zone (see, for example, \citet{thompsonetal03},
and references therein). Observation of the meridional
flow is a more difficult task. We now know its approximate shape near
the solar surface. The temporal analysis of these profiles support
the use of the kinematic approximation for the solar dynamo,
since the temporal variations in these profiles are small. From the
observational point of view we can argue that the changes in these
plasma movements due to the back reaction of magnetic fields are
either small, or replenished in a time-scale much shorter than the
solar cycle.

In the present calculation we incorporate the velocity profiles
as prescribed in eqs. (4) and (5) of \citet{dc99}, which resemble the
helioseismology results regarding the differential rotation, and
assume one meridional flow cell per meridional quadrant, which is a
likely assumption when both the observed all-latitude poleward
flux and the mass conservation law are considered together. We
notice that, despite several attempts have been made in
flux-transport dynamo models to include more than one convective cell
\citep{bonnetal06,jouve07}, no inferred magnetic field distribution
agrees better with the observations than the one that is obtained
by considering only one cell pattern \citep{diketal04,gdp07b}.
Note however, that the depth of penetration, as well as the
magnitude of the flow in the deeper layers are still uncertain
in these models. In fact, nothing is known from observations
regarding the structure and the amplitude of the flow pattern inside 
the convection zone, except only very near the surface.

\begin{figure}[hbt]
\epsscale{1.0}
\plotone{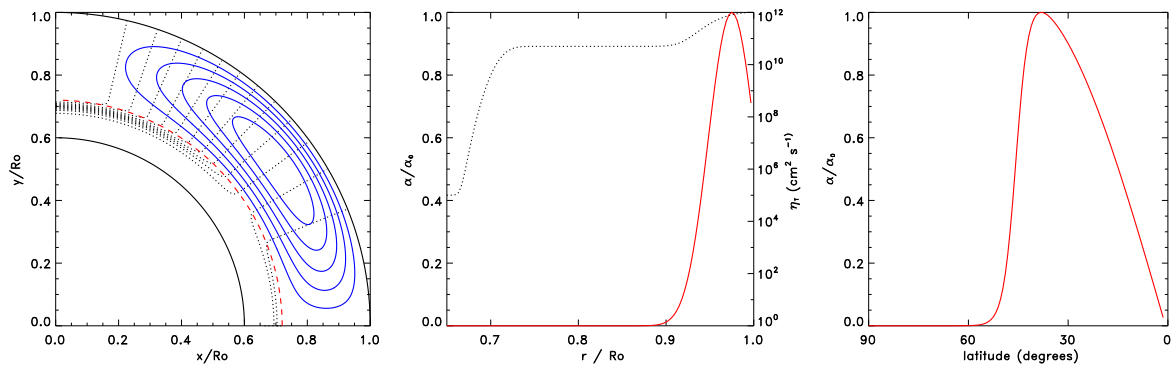}
\caption{
Profiles of the main ingredients of the solar dynamo. The
left panel shows the contours of iso-rotation (dotted lines) together
with the meridional flow streamlines (solid contours), the dashed
line at   $r$$=$$0.715 \sr$
indicates the center of the overshoot region. The middle panel
shows the radial variation of the BL $\alpha$ term (solid line)
and the magnetic diffusivity, $\eta_{T}$ (dotted line). The values
for $\alpha$ are normalized
to its maximum value $\alpha_0$, and the values of $\eta_{T}$ are at
the right axis. The right panel shows the latitudinal profile of the
BL $\alpha$ term.
}
\label{fig1}
\end{figure}

Both the contours of the iso-rotation and the streamlines of the
meridional flow are shown in the left panel of Fig. 1. For all
simulations presented below we consider a meridional flow amplitude
of $u_0/2$$=$$15$ m s$^{-1}$ and a depth of penetration $r_p$$=$$0.7\sr$.
Note that $u_0/2$ gives the maximum flow-speed (i.e. 15 ${\rm m}\,
{\rm s}^{-1}$) for a maximum $u_0=30 \,{\rm m}\, {\rm s}^{-1}$ for
the mathematical prescription of meridional circulation we are using.
In this calculation, we consider that the tachocline is centered at 
$0.7\sr$ with a thickness $\omega$$=$$0.03\sr$ (see e. g. \citet{gdp07a}).

\subsection{Diffusivity profile}

The magnetic diffusivity is another important ingredient in 
dynamo models, but we know very little about the amplitude and
profile of this ingredient in the solar interior. At the surface, it
should be of the same order as the supergranular diffusion, and
at the radiative layer and beneath it should attain molecular values.
However we do not have a good idea what value it should have in the
bulk of the convection zone. Recently the turbulent diffusivity
for large-scale magnetic fields has been estimated using the so-called
test field method for isotropic turbulence \citep{sbs08} and
convection \citep{kkb09}. According to these studies, the turbulent
diffusion is of the same order of magnitude as the first order smoothing
estimate. This is in agreement with mixing length arguments for the
kinematic regime in the range of currently accessible Reynolds number.

While the calculation by \citet{kkb09} is a substantial advance and
the best available today, given the limitation of computer powers,
it is still far from solar-like conditions in terms of Reynolds
number and the stratification. Inclusion of more realistic solar 
conditions as well as the ${\bf j}\times {\bf B}$ back-reaction
that allows the formation of intense flux tubes may reduce the effective
diffusivity in the bulk of convection zone. Furthermore, previous
flux-transport dynamo studies have shown that the magnetic diffusivity
is required to be one order of magnitude lower in the convection zone 
than at the surface in order for the dynamo to operate in the
advection-dominated regime. So, in this
work we use a diffusivity profile that has been used previously in 
several works (e.g., \citet{diketal02, gdp07a}), as follows:

\begin{equation}\label{eq4}
\eta_T(r) = \eta_{rz} + \frac{\eta_{cz}}{2}\left[1+erf
  \left(\frac{r-r_{c}}{d_1}\right)\right]
+\frac{\eta_{s}}{2}\left[1+erf
  \left(\frac{r-r_{c1}}{d_2}\right)\right]\quad ,
\end{equation}

\noindent where $\eta_{rz}$$=$$10^5$  cm$^2$ s$^{-1}$,
$\eta_{cz}$$=$$5\times10^{10}$  cm$^2$ s$^{-1}$ and
$\eta_s$$=$$10^{12}$ cm$^2$ s$^{-1}$ correspond to the values of the
diffusivity at
the radiative, convective and near-surface layers, respectively. The
transition from the radiative to the convective layers is
located at $r_{rc}$$=$$0.715\sr$ (the overshoot interface), with
$d_1$$=$$0.015\sr$, and the transition from the turbulent convective
zone to the (sub-surface) supergranular diffusion layer is at
$r_{rc}$$=$$0.96\sr$, with $d_1$$=$$0.03\sr$ (see the dotted line in
the middle panel of Fig. 1).

\subsection{Formulation of dynamo equations with $\eta$-quenching}

As mentioned earlier, the main goal of this work is to study the
quenching of the turbulent diffusivity due to the presence of strong
magnetic fields. For this aim we will assume that it will affect only
the toroidal fields, and replace $\eta_T$ in the equation for $B$
(eq. \ref{eq3}) by $\eta$:

\begin{equation}\label{eq5}
\eta = \frac{\eta_T}{1 + (B/B_q)^2}\quad,
\end{equation}

\noindent
which is the same algebraic form for the $\eta$-quenching used
by \cite{gr05}. In the equation above, $B_q$ is the value of the
magnetic field at which $\eta_T$ begins to be quenched.
In principle, the poloidal fields also could contribute to
the diffusivity quenching. However, here we focus only on the 
influence of toroidal fields in the saturation mechanism of the
the turbulent diffusivity for two reasons: (1) in most
of the $\alpha$-$\Omega$ solar dynamos, the dynamo-generated toroidal 
fields are about a thousand times stronger than the poloidal fields;
(2) the amplitude of poloidal fields generated in a Babcock-Leighton 
flux-transport dynamo is of the order of a few hundred Gauss, much
below the value of the lowest quenching field strength selected for
the study in this paper. 
In a more realistic situation, the diffusivity is a tensor and
its quenching is bound to have some effect due to the presence of 
poloidal fields also, and must be explored in future. 

Before we re-derive the dynamo equations with the inclusion of
$\eta$-quenching, we briefly discuss the issue regarding the choice of
$\eta$-quenching formula. Analytical studies considering 3D turbulence
often yield a quenching formula proportional to $|B|$ rather than
$B^2$ \citep{kpr94,rk01}. 

In turbulent forced MHD simulations, \cite{yetal03} found that the
suppression of the magnetic diffusivity follows the form:  
$\eta$$\simeq$$\eta_{T0}/(1+a({{\bf B}/B_{eq}})^2)$, where the value 
of $a$ depends on the geometry of the
initial magnetic field (i.e. whether it is helical or not). 
K\"apyl\"a and colleagues (see $http:\/\/arxiv.org\/abs\/0810.2298$)
have also recently performed numerical simulation where turbulent
diffusivity is quenched. However, their results are not able to make 
a clear conclusion about the functional form of quenching to be 
proportional to $|B|$ or $B^2$. In the
present case, we will adopt a similar formulation, but with $B_q$ as a
free parameter of the model. The results of
\cite{yetal03} indicate that in the solar convection zone
$\eta_{T}$ is quenched approximately as in eq. (\ref{eq5}) for
fiducial choices of the value of $B_q$.

We note that, with the quenching incorporated, $\eta$ is not only a
function of $r$, but  also depends on $B(r,\theta,t)$, and hence
$\eta(r,\theta,t)$ is a function of $r$, $\theta$ and $t$. Due to
the additional dependence of $\eta(r,\theta,t)$ on $\theta$ and $t$,
the $\theta$ component of $\nabla \eta \times (\nabla \times B
\hat{e}_{\phi})$ gives rise to two new terms in the induction
equation.

Thus the Equation (3) becomes:

\begin{eqnarray}
\frac{\partial B}{\partial t}+\frac{1}{r}\left[\frac{\partial}{\partial
    r}[r u_r B]+\frac{\partial}{\partial
    \theta}[u_{\theta} B]\right]\label{eq3.1}=s({\bf
  B_p} \cdot \nabla) \Omega\\\nonumber
-\left[\frac{\partial \eta}{\partial r} \left(\frac{\partial
    B}{\partial r} - \frac{B}{r}\right) + \frac{1}{r^2}\frac{\partial
    \eta}{\partial \theta} \left(B \cot \theta + \frac{\partial
    B}{\partial \theta} \right) \right]_{\phi} \\\nonumber
+\eta \left(\nabla^2-\frac{1}{s^2}\right)B \nonumber \quad,
\end{eqnarray}

\subsection{Babcock-Leighton $\alpha$ effect}

The source of poloidal fields (eq. \ref{eq2}) in a Babcock-Leighton
dynamo is the decay of the tilted bipolar magnetic regions (BMR's),
which form a net surface dipole moment that drift towards the poles
and eventually cause the reversal of the polar fields. The place
where the magnetic flux tubes, which are responsible for the
formation of the BMR's, develop is uncertain, but due to various
reasons we can assume that the flux tubes are formed at or below the
base of the convection zone. There exists a vast literature on the
topic of rising flux tube simulations that produce the tilt and emergence
pattern at the solar surface which are in good agreement with
observations (see, for example, \citet{dh93, ffm94}). The 
combination of low magnetic
diffusivity and helioseismically obtained differential rotation
helps the amplification of toroidal fields there.

Furthermore, the rising flux tube simulations indicate that their
eruption latitude and the tilt acquired during their buoyant rise
through the convection zone fit best with surface observations
if flux tubes with magnitudes between $5 \times 10^4$ and $10^5$ G
are produced at the bottom of the convection zone. \footnote{See,
however, a discussion of alternative possibilities in
\cite{diketal02, bran05,gdp08}}.

Estimating the Babcock-Leighton $\alpha$-effect by computing the
buoyant eruption of magnetic flux tubes followed by their decay is
beyond the scope of this paper. So, in order to capture the
properties of  the BMR's described above, we simply use the
following Babcock-Leighton $\alpha$ effect profile
\citep{diketal04}:

\begin{equation}\label{eq6}
S_1(r,\theta,B) =
\alpha(r,\theta) f_Q(\overline{B_{r_c}}) \overline{B_{r_c}}
\quad .
\end{equation}

\noindent This term is non-local in $B$ (see below), $\overline{B_{r_c}}$
being the radial average of $B$ between $r$$=$$0.7\sr$ and $r$$=$$0.72\sr$;
this is a simple way to erupt toroidal magnetic flux from the bottom
of the convection zone to the place where the $\alpha$-effect is operating.
We incorporate the following radial and latitudinal dependence of the
$\alpha$-effect:

\begin{eqnarray}\label{eq7}
\alpha(r,\theta)&=&\alpha_0
\frac{1}{4}\left[1+\mathrm{erf}\left(\frac{r-r_2}{d_2}\right)\right]
\left[1-\mathrm{erf}\left(\frac{r-r_3}{d_3}\right)\right] \\\nonumber
&\times& \sin\theta
\cos\theta\left[\frac{1}{1+e^{\gamma_1\left(\pi/4-\theta\right)}}\right]
\quad,
\end{eqnarray}

\noindent where $r_2$$=$$0.95\sr$, $r_3$$=$$\sr$,
$d_2$$=$$d_3$$=$$0.01\sr$, and $\gamma_1$$=$$30$. The amplitude of the
poloidal source is determined by $\alpha_0$, for which we assume a
fixed value of $50$ cm s$^{-1}$ for the first set of simulations
below (from now onwards we will explicitly quote the value of $\alpha_0$
only when we are using a value different from this). As can be seen in
Fig. 1, the $\alpha$-effect is distributed  mainly at the low latitudes
peaking around $45^{\circ}$; it is also concentrated above $0.95\sr$
since it is expected that the main poloidal field component is formed
at the sunspot latitudes near the surface \citep{wangetal89,wang91}.

The quenching of the poloidal source term (the second term of eq.
\ref{eq6}) is given by:

\begin{equation}\label{eq8}
f_Q(B) =
\left(1+\left[\frac{\overline{B_{r_c}}(\theta)}{B_0}\right]^2\right)^{-1}
\quad.
\end{equation}

This term has the same algebraic form as the $\alpha$-quenching
term used in turbulent mean field dynamo models. 
Its function here is to saturate the growing of
the poloidal fields when the toroidal field at the base of the
convection zone is around $B_0$$=$$10^4$ G. Thus the model will
produce toroidal magnetic fields in the expected range, as explained
above. The amount of poloidal field that will be produced from this
toroidal field is determined by $\alpha_0$.
We note that this is not the unique way of capturing the physics
of buoyantly erupted flux tubes. \citet{nandy} have replaced this
non-local quenching term by a different buoyancy mechanism term
parameterized from simulations of flux tubes.

Although the equations (\ref{eq5}) and (\ref{eq8}) have the same
functional form, they operate in opposite ways. We will discuss
that in detail in \S4. On the one hand, smaller values of $B_q$ in eq.
(\ref{eq5}) will produce lower values of $\eta$ which will, in turn,
help to increase the amplitude of the toroidal component, $B$. On
the other hand, lower values of $B_0$ in eq. (\ref{eq8}) will limit
the growth of $B$ up to values around $B_0$. Both these terms are
the sources of non-linearity in the model.

\section{SOLUTION METHOD, BOUNDARY AND INITIAL CONDITIONS}

We solve equations (\ref{eq2}), (\ref{eq3}) and (\ref{eq5}) for
$A$ and $B$ and $\eta$, with the coordinates $r$ and $\theta$
covering the spatial range $0.6 \sr$$\le$$r$$\le$$\sr$ and
$0$$\le$$\theta$$\le$$\pi/2$, which spans from the outer most 
part of the radiative zone, through the overshoot layer (where  the
tachocline is located) and the entire convection zone, up to the
surface, in the northern hemisphere. We have used a second order
finite difference scheme for the spatial discretization; the
Lax-Wendroff method for the first order derivatives, and centered
finite difference for the second order derivatives. The temporal
evolution is solved with the ADI semi-implicit method
(see \citet{gm04,dc99}, for details). In eq. (\ref{eq5}), the
dependence of $\eta$ with time is implicit, thus we update the
value of $\eta$ each half-time step with the previous values of $B$
and then we calculate the derivatives of $\eta$ and use these values
to solve the equation for $B$.

The boundary conditions are: $A=0$ and $B=0$ at the north pole
($\theta=0$); and at the equator $B=0$, but $A$ is coupled
with the southern hemisphere in such a way as to ensure
antisymmetric magnetic fields about the equator, so we demand
${\partial \over \partial \theta} \left(r \sin\theta A \right)=0$. 
At the bottom radial boundary we use $A={\partial \over \partial r}(rB)=0$, 
and finally, at the upper radial boundary $r=\sr$, we consider a 
potential field boundary, i.e., $B=0$ and $A$ coupled to an external 
vacuum field $(\nabla^2 + 1/s^2) A=0$. A complete description of the 
boundary conditions and their numerical implementation can be found 
in \cite{dc94,dc99}.

Since our intention is to explore how the $\eta$-quenching affects a
magnetic field that is well organized both in space and time, a
plausible choice for an initial condition for the simulation is to
start with a fully relaxed solution of the same dynamo without
$\eta$-quenching. So we first obtain a fully converged solution by
initializing the system with $A=\sin\theta / r^2$  if
$r \ge 0.715$, $A=0$ otherwise, and $B=0$, and allow it to
evolve until $10^4$ yr with the $\eta$-quenching in the Equation (6)
turned off.

\begin{figure}[hbt]
\epsscale{1.0}
\plotone{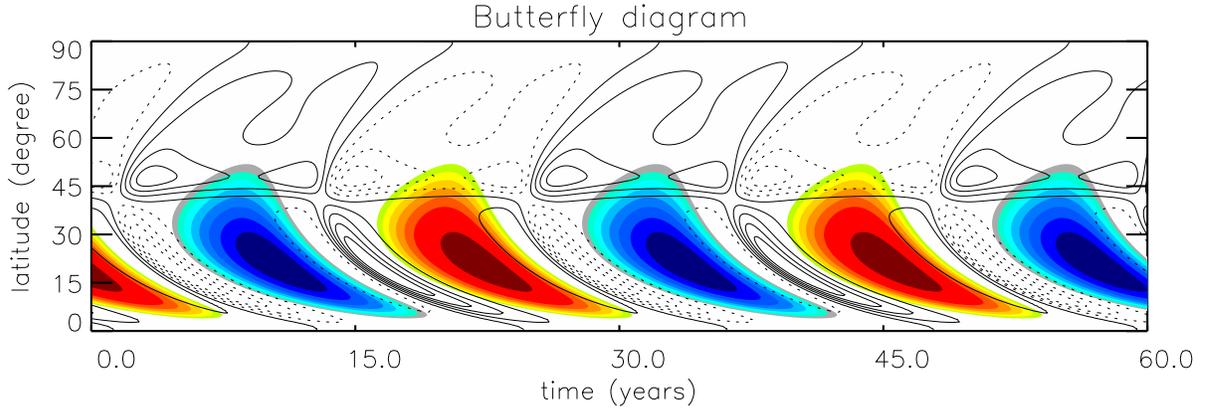}
\caption{
Butterfly diagram for the relaxed solution of the reference
model. The dark (blue) contours represent positive toroidal fields
and the light (red) contours represent the negative toroidal
fields. The plotted
values correspond to a radial average between $0.7\sr$ and $0.72\sr$.
Only contours for fields above $1.2\times 10^4$
G are plotted. The continuous and dashed lines represent the
positive and negative radial fields at the surface,
respectively.
}
\label{fig2}
\end{figure}

Figure \ref{fig2} shows the time-latitude, butterfly diagram for the 
last $60$ years of evolution of our reference simulation. It shows 
the gray scale (color) contours of the radially averaged toroidal 
magnetic field $\overline{B_{r_c}}$ (in log-scale) with values above 
$12$ kG, together with the contours of the radial field $B_r$ at the 
surface. It can be seen that the maximum amplitudes of the toroidal field
are located below $45\degree$, and satisfy the $\pi/2$ phase-lag observed
with respect to the radial field. The maximum value of the toroidal
magnetic field $\overline{B_{r_c}}$ is $\sim22.9$ kG, the maximum
value of the radial field is $86.6$ G, and the period of the entire 
cycle is $21.2$ yr.

\clearpage

\begin{figure}[hbt]
\epsscale{0.3}
\plotone{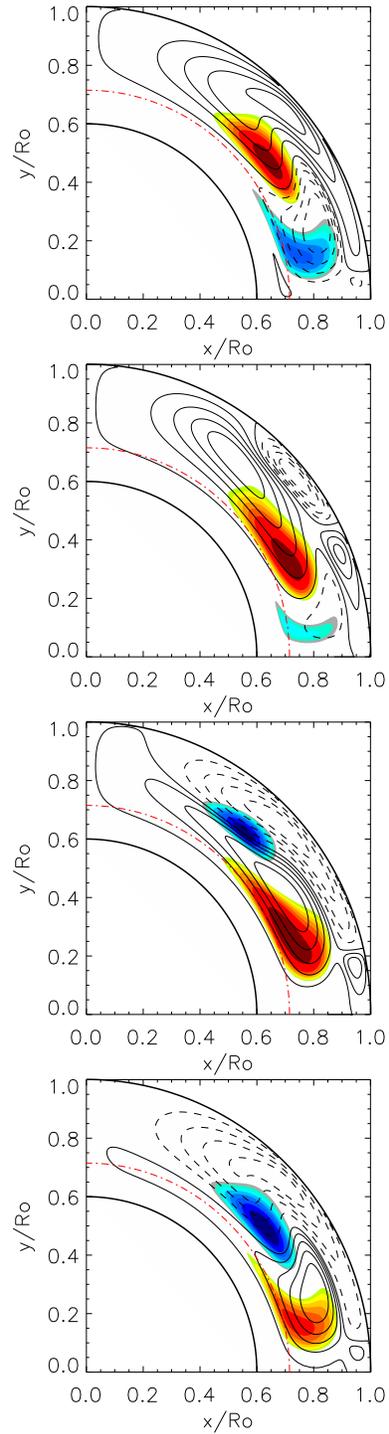}
\caption{
Snapshots of the relaxed solution of our reference model at
$0$, $T/8$, $T/4$ and $T/2$, where T is the full period of the
cycle. The gray scale (color) contours follow the same description
as in Fig. \ref{fig2}, but in this case the line contours correspond
to the total poloidal field.
}
\label{fig3}
\end{figure}

\clearpage

The four frames of Fig \ref{fig3} (a-d) show the same
temporal evolution in the meridional cut. We see that the
strongest toroidal fields begin to form at mid-latitudes inside the
convection zone ($r$$\sim$$0.8\sr$). The toroidal field penetrates
slightly in the overshoot layer only at lower latitudes, but it is
not substantially amplified there by the radial shear. This happens
due to two reasons: first, since the meridional flow is not allowed in
this model to go deep inside the overshoot layer, not enough poloidal 
fields (which are actually the source for the toroidal fields) can 
reach the radial shear layer; second, the radial component of the 
poloidal field is much weaker there than the latitudinal component
\citep{gdp07a}. Fig \ref{fig3} also reveals that the positive and
negative poloidal magnetic fields (continuous and dashed lines,
respectively) are produced at mid-latitudes at the surface and
then migrate poleward following the plasma flow.

We use the converged solution without the $\eta$-quenching shown in
Figures \ref{fig2} and \ref{fig3} as our initial ($t=0$) configuration
for our simulations with $\eta$-quenching and run the model
for 200 years more. In \S4, we present our simulation results with
$\eta$-quenching.

\clearpage

\section{Results}

\subsection{The effect of $\alpha$-quenching term in Babcock-Leighton
  dynamos}

Generally $\alpha$-quenching is applied in most of the large-scale,
mean-field dynamo models and the basic results are known. In a
kinematic dynamo, the maximum value that the toroidal field can
reach depends on the amount of poloidal field being generated by the
Babcock-Leighton $\alpha$-effect, and the later depends on the
values of $\alpha_0$ and $B_0$ in Equations (\ref{eq7}) and (\ref{eq8}). 
In order to explore the influence of $\alpha$-quenching in 
more detail and to compare this influence with that obtained by
implementing the $\eta$-quenching, we study how the maximum toroidal
fields produced at different latitudes vary with the quenching field
strength. The value of $\alpha_0$ defines the non-dimensional number
$C_{\alpha}$$=$$\alpha_0 \sr / \eta_{cz}$$=$$70$ for
$\alpha_0$$=$$50$ cm s$^{-1}$ and $\eta_{cz}$$=$$5 \times 10^{10}$
cm$^2$ s$^{-1}$, this value remains constant during all the
simulations shown below. This guarantees that the dynamo efficiency
$C_{\Omega}C_{\alpha}$ is always the same. Then, we change the value
of $B_0$ in eq. (\ref{eq8}) between $5\times10^2$ G and $5
\times10^5$ G and measure the maximum value that the toroidal field
reaches at the numerical domain during one-half period. Intuitively
we expect that the maximum value of $B_{max}$ should be larger
for larger $B_0$.

Fig. 4 presents $B_{max}$ as function of $B_0$ for two different
radii $r$$=$$0.7\sr$ and $r$$=$$0.8\sr$, and three different
latitudes, $10\degree$ (continuous line), $45\degree$ (dashed line)
and $80\degree$ (dotted line). Figure 4 immediately reveals that
maximum toroidal fields generated at different latitudes and
different depths vary near linearly with the quenching field
strength; the higher the quenching field strength the higher the
dynamo-generated toroidal field. This means that the nonlinearity
due to $\alpha$-quenching is so weak even up to a quenching field
strength of $10^5$ Gauss that the dynamo behaves virtually as if it 
is operating in the linear regime.

We also see that for lower latitudes ($<45 \degree$), the maximum
toroidal field is of the same order both in the convection zone
($r$$=$$0.8\sr$) and at the tachocline ($r$$=$$0.7\sr$). This is
because the largest toroidal fields in a Babcock-Leighton
flux-transport dynamo are produced mainly by the latitudinal shear
working on the latitudinal poloidal fields rather than by the action
of tachocline radial shear on radial fields. This reinforces the
result obtained previously by some authors \citep{rempel06, ddg06,
gdp07a}. With no advective transport below the base of the
convection zone, very little poloidal field diffuses down there, and
due to the thinness of the tachocline, an even smaller radial
component of those poloidal fields is available to be sheared there.
However, the situation can be different if the magnetic fields can
be transported downwards by overshooting or magnetic pumping at the
lower latitudes and then acquire further amplification at the radial
shear layer \citep{gdp08}.

\clearpage

\begin{figure}[hbt]
\epsscale{1.0}
\plotone{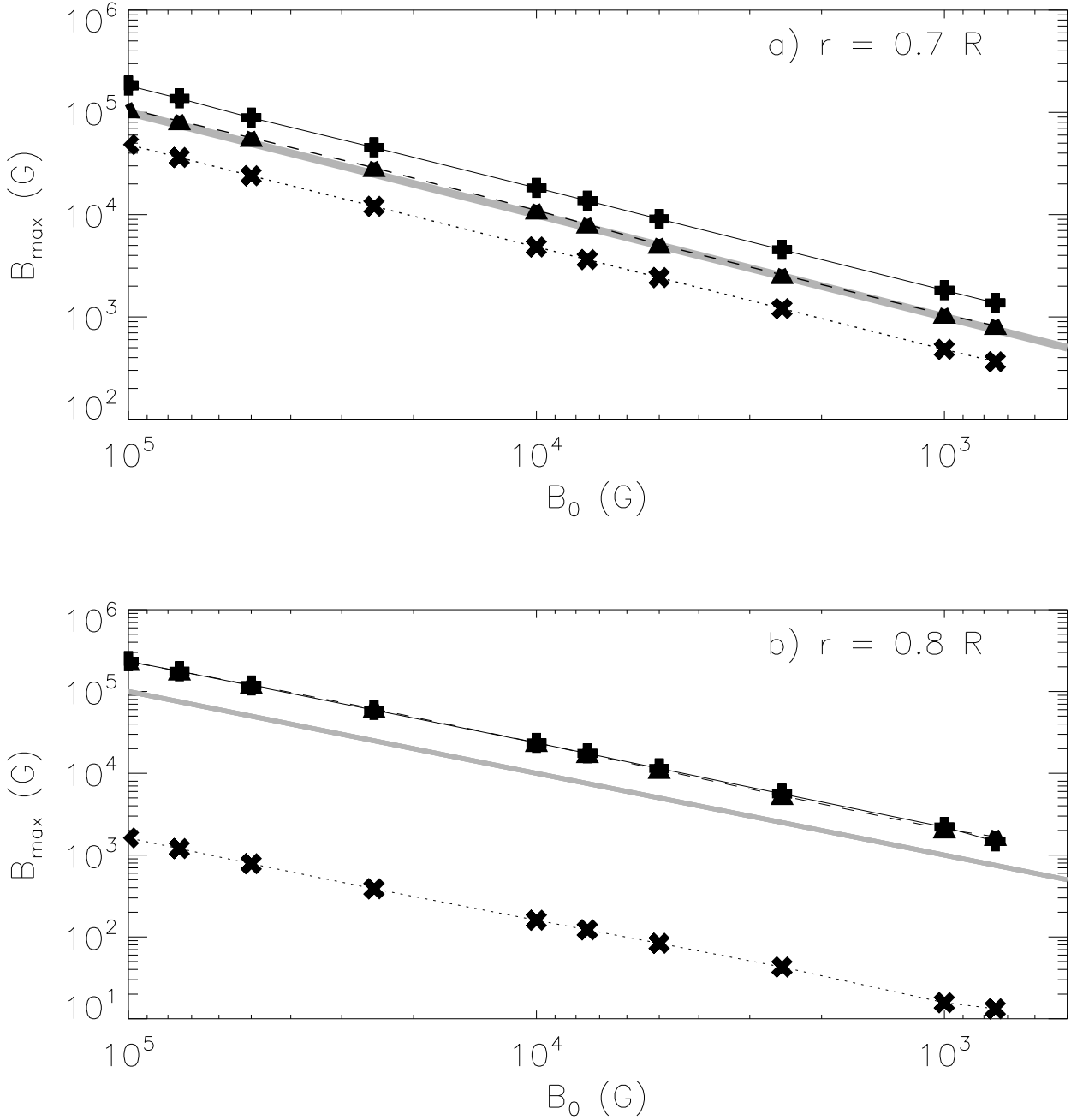}
\caption{
Maximum toroidal magnetic field as function of $B_0$ at
  $r$$=$$0.7\sr$ and $r$$=$$0.8\sr$ (see the
  $\alpha$ quenching function, eq. (\ref{eq8})) for
  three different latitudes, $10\degree$ (continuous line, cross
  symbol), $45\degree$ (dashed line, triangle symbol) and $80\degree$
  (dotted line, $X$ symbol). In the bottom panel, the continuous and
  dashed lines are almost coincident. In each panel, a $B_{\rm max}=B_0$
  curve has been superimposed to show the linear proportionality
  of maximum toroidal fields produced by a certain $\alpha$-quenching field
  strength.
}
\label{fig4}
\end{figure}

\clearpage

\subsection{The effects of $\eta$-quenching in magnetic field evolution}

It has already been noted in the context of one-dimensional $\alpha\Omega$
dynamo, two-dimensional $\alpha^2 \Omega$ dynamo and interface dynamo that,
the quenching of the magnetic diffusivity due to the back-reaction of
magnetic fields is a possible mechanism for further field amplification.
In this subsection, we first explore in detail the evolution of
magnetic fields in a diffusively-quenched Babcock-Leighton flux-transport
dynamo, starting from the 2D reference-state solution described in \S3.
Subsequently we will also present the quantitative estimation of field
amplification and the change in cycle period due to $\eta$-quenching.

In the case of the $\alpha$-quenching study, we ran our
simulation for $5\times10^2 {\rm G} \le B_0 \le 5\times10^5 {\rm
G}$. We consider same range for $B_q$ for the $\eta$-quenching study:
$5\times10^2 {\rm G} \le B_q \le 5\times10^5 {\rm G}$,
and keep all the other parameters the same as in the reference model
described in \S3. Using a fully converged solution of the reference
model as the initial condition, we switched on the $\eta$-quenching
and ran the simulation for $200$ years. At this time, the system
reaches a new steady state with cyclic variations of $A$, $B$ and
$\eta$.

It is not surprising that, the smaller the $B_q$, the faster the
$\eta_T$ is quenched, leading to a significant increase in
toroidal field. Figures 5 and 6  show the temporal evolution of both,
the toroidal and poloidal fields in the pole-to-equator meridional-cut
for four successive times within half a cycle (left panels) for two
different representative values of $B_q$ ($10^4$ G and $10^3$ G,
respectively). These are the solutions after 200 years' evolution.
Right panel shows the change of $\eta$ due to the quenching action,
for two different latitudes, $10\degree$ (solid line) and $45\degree$
(dashed line). The non-quenched profile has been plotted also for
comparison (red dotted line).

The common features in both cases (see Figures 5 and 6) are that
the model exhibits a decrease in the diffusivity at the places
where the toroidal field acquires considerable amplitude.
This suppression can be as large as three orders of magnitude, as
we see in Figure 6 for $B_q=10^3$ G, but this suppression of $\eta$
is not uniform -- neither along the radial direction nor in
latitude. Comparing the left and right panels in each of Figures 5
and 6, we see that the peaks and valleys are anti-correlated
with the spatial distribution of the toroidal field amplitudes.
Since we are presenting here the converged solution, this change
in $\eta$ repeats in successive cycles.

\clearpage

\begin{figure}[hbt]
\epsscale{0.6}
\plotone{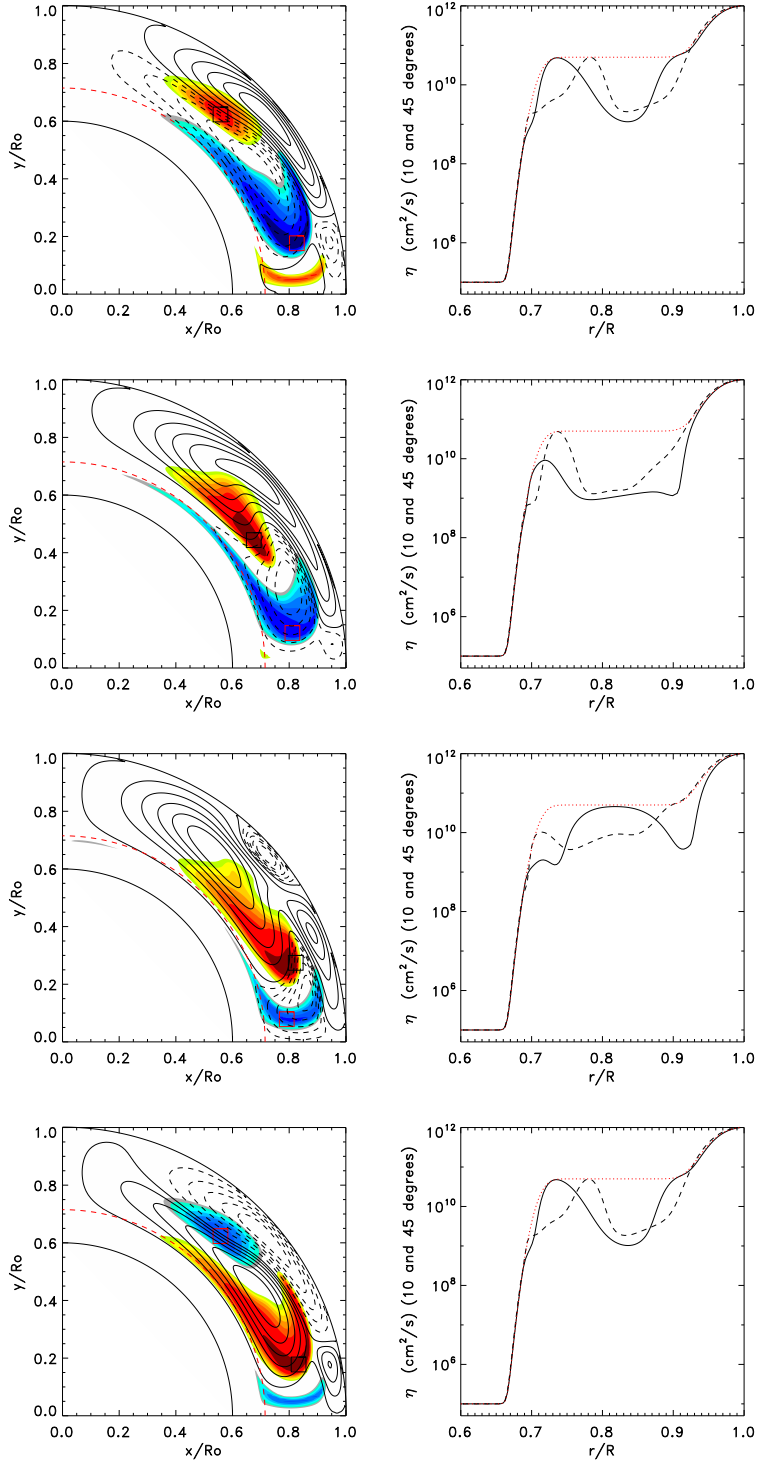}
\caption{
Left panel: Snapshots for 4 different times within a half
  period of the relaxed model for $B_q$$=$$10^4$ G. Right panel:
  Turbulent diffusivity without quenching (red dotted line), for
  $10\degree$ (continuous black line) and $45\degree$ (dashed blue
  line).
}
\label{fig9}
\end{figure}

\clearpage

\begin{figure}[hbt]
\epsscale{0.6}
\plotone{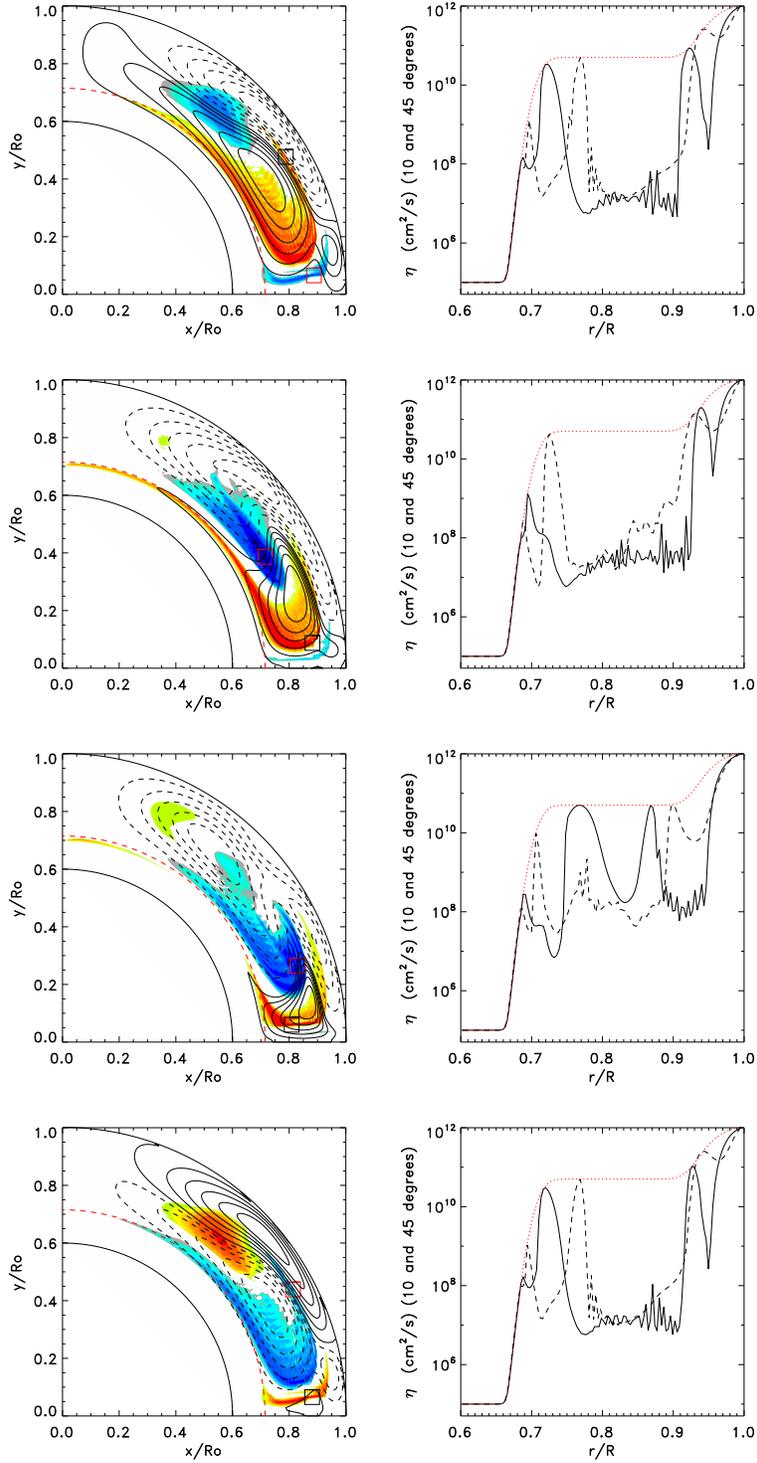}
\caption{
The same as in Fig. \ref{fig9}, but for $B_q$$=$$10^3$ G.
}
\label{fig8}
\end{figure}

\clearpage

In both cases, strong gradients in diffusivity in latitude as well
as in depth occur, leading to the more efficient field amplification
in those regions and hence, the formation of small regions of 
concentrated toroidal magnetic fields. Comparing Figures 5 and 6 with 
Figure 3, we can see the enhancement in magnetic fields even in places
where a run without quenching does not exhibit strong toroidal fields.

It is not difficult to understand why the diffusivity quenching
leads to small regions of flux concentrations in the computation
domain. The average diffusion time of the fields at each portion
of the domain where $\eta$ is strongly suppressed is larger than
the neighbouring domains where the quenching is less effective.
The lifetime of the toroidal fields is several years larger there
than in the regions where the quenching is not as effective. Thus
the former can undergo a prolonged amplification by the $\nabla \Omega$
terms and can reach larger values than the neighbouring regions
that have larger $\eta$ .

An interesting feature we note in Figure 6 is the formation of small
scale magnetic patterns at the overshoot tachocline (see in the left
panels). This is associated with a variation in $\eta$ in finer
spatial scale compared to what we see in Figure 5 as a function of
depth. This is a consequence of the non-linear coupling between
$B_q$ and $\eta$. The smaller the $B_q$, the narrower the $\eta$
profile, and the more segregation of the magnetic fields is produced
\footnote{We have tested  the model calculation using a larger grid
resolution, such as $200\times200$ grid points and the results
remain unchanged}. Due to large suppression in $\eta$ at the
overshoot tachocline regions, the spot-producing toroidal flux 
remains more frozen there, particularly in
the case of Figure 6. In the meantime, two competing processes are
going on -- the prolonged shearing by the differential rotation, and
the equatorward advection due to meridional flow. Since this advection
must work against the diffusion of the fields, we can think of it as
``dragging'' or ``pulling'' the fields along, at some net speed that is
smaller than the meridional flow there. If the equatorward advective 
drag partially wins at certain portions of these fields, those 
portions get torn out from the large-scale part of the fields,
and locally reconnect. This happens predominantly on the equatorial
side of the large-scale fields, and so more and more fragmentation 
of field takes place, leading to formation of small-scale
structures.

\subsection{Influence of $\eta$-quenching on field amplification}

In order to quantify how effective the $\eta$-quenching is in
producing strong toroidal fields, we have plotted in Fig. 7 the
maximum value of the dynamo-generated toroidal field as a function
of $B_q$. As in Fig. 4, we show the results at three different
latitudes: $10\degree$ (solid line), $45\degree$ (dashed line) and
$80\degree$ (dotted line), for two different radii, namely for
$0.7\sr$, the center of the tachocline, and $0.8\sr$, the lower
convection zone.

At the tachocline (upper panel of Fig. 7), the curves show an
interesting behaviour: at latitudes close to the poles (dotted
line), the toroidal field increases with decreasing $B_q$ (i.e.,
with increasing quenching). The amplification factor could
be up to $\sim 2.5$. We can understand this, because the poloidal
fields carried down to the tachocline by the meridional flow,
undergo prolonged shearing by the strong radial differential rotation
there. However, we obtain an apparently counter-intuitive result
that for lower latitudes (solid lines), there is a decrease of $B$
with the increased quenching in $\eta$ (decrease in $B_q$). We see
a little decrease of $B$ at mid-latitudes also as $B_q$ decreases,
but the effect is not so pronounced, and appears only for very
efficient quenching factors. For the fields at $10\degree$, the
factor of decrease could be as large as $\sim 2.5$.

At the center of the convection zone (bottom panel of Fig. 7), the
results are different; the magnetic fields for both low and mid-latitudes 
have the same increase, by a factor as large as  $\sim 2$
with respect to the case with no quenching. With $B_q=5 \times 10^3$
G the toroidal fields can reach values above $10^5$G. In the middle
of the convection zone there is no radial shear. It is the
latitudinal shear that works on the poloidal fields, so we see
similar amplification for mid-latitude and low-latitude fields.
If the flux tubes formed from these strong toroidal fields produced
at the middle of the convection zone due to strong $\eta$-quenching,
rise to the surface, their orientation may not agree with Joy's law. 
So we do not know whether such a strong $\eta$-quenching is working 
in reality, or some other processes are inhibiting the formation of
flux tubes in the middle of the convection zone.

\clearpage

\begin{figure}[hbt]
\epsscale{1.0}
\plotone{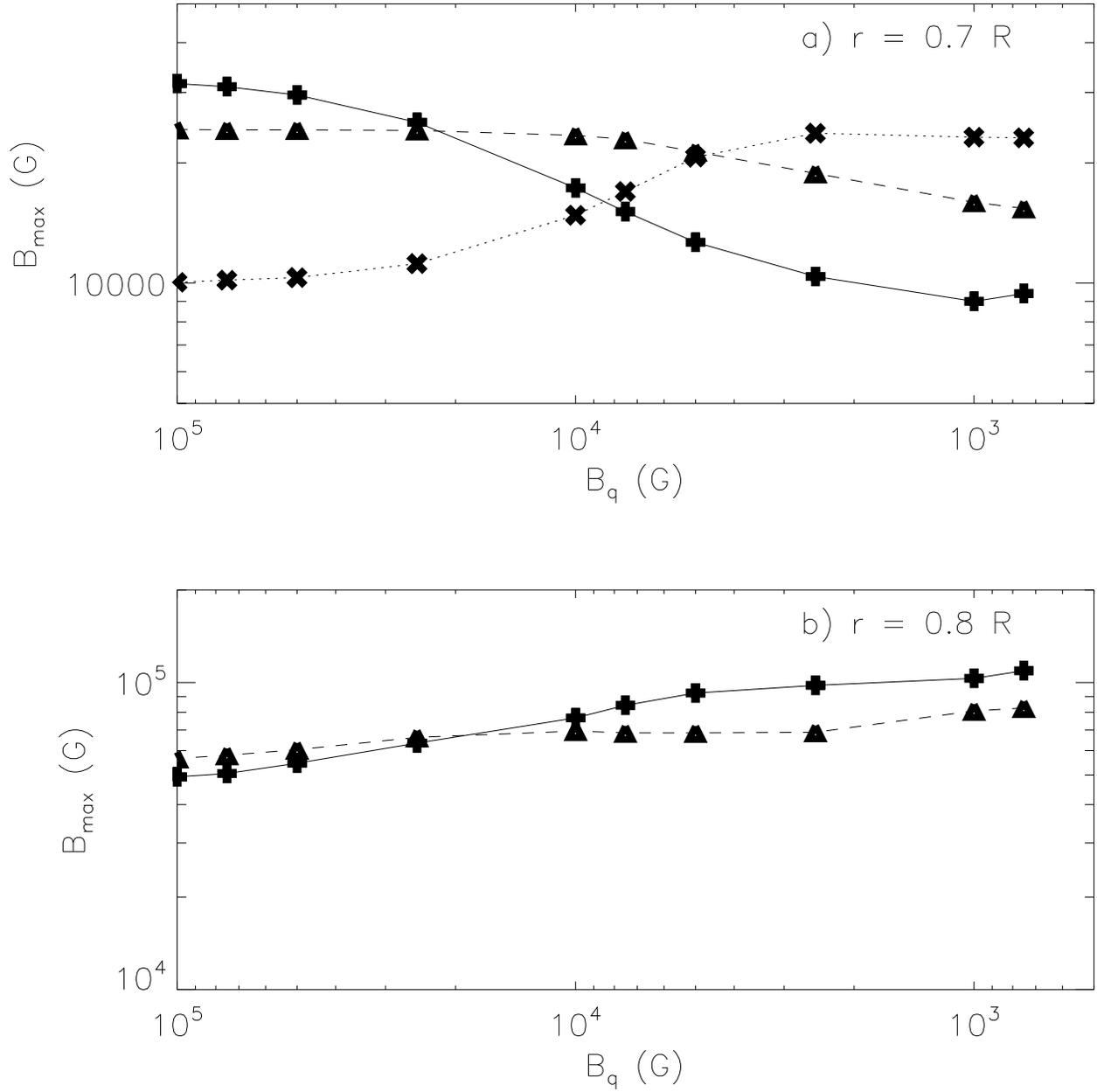}
\caption{
Maximum toroidal magnetic field as function of $B_q$ (see the
  eq. (\ref{eq5}) for three different latitudes, $10\degree$
  (continuous line, cross symbol), $45\degree$ (dashed line, triangle
  symbol) and $80\degree$ (dotted line, $X$ symbol). At the
  bottom panel, only the results for $10\degree$ and $45\degree$ are
  shown, the maximum values of $B$ for $80\degree$ are smaller than
  $10^4$ G.
}
\label{fig5}
\end{figure}

\clearpage

\begin{figure}[hbt]
\epsscale{1.0}
\plotone{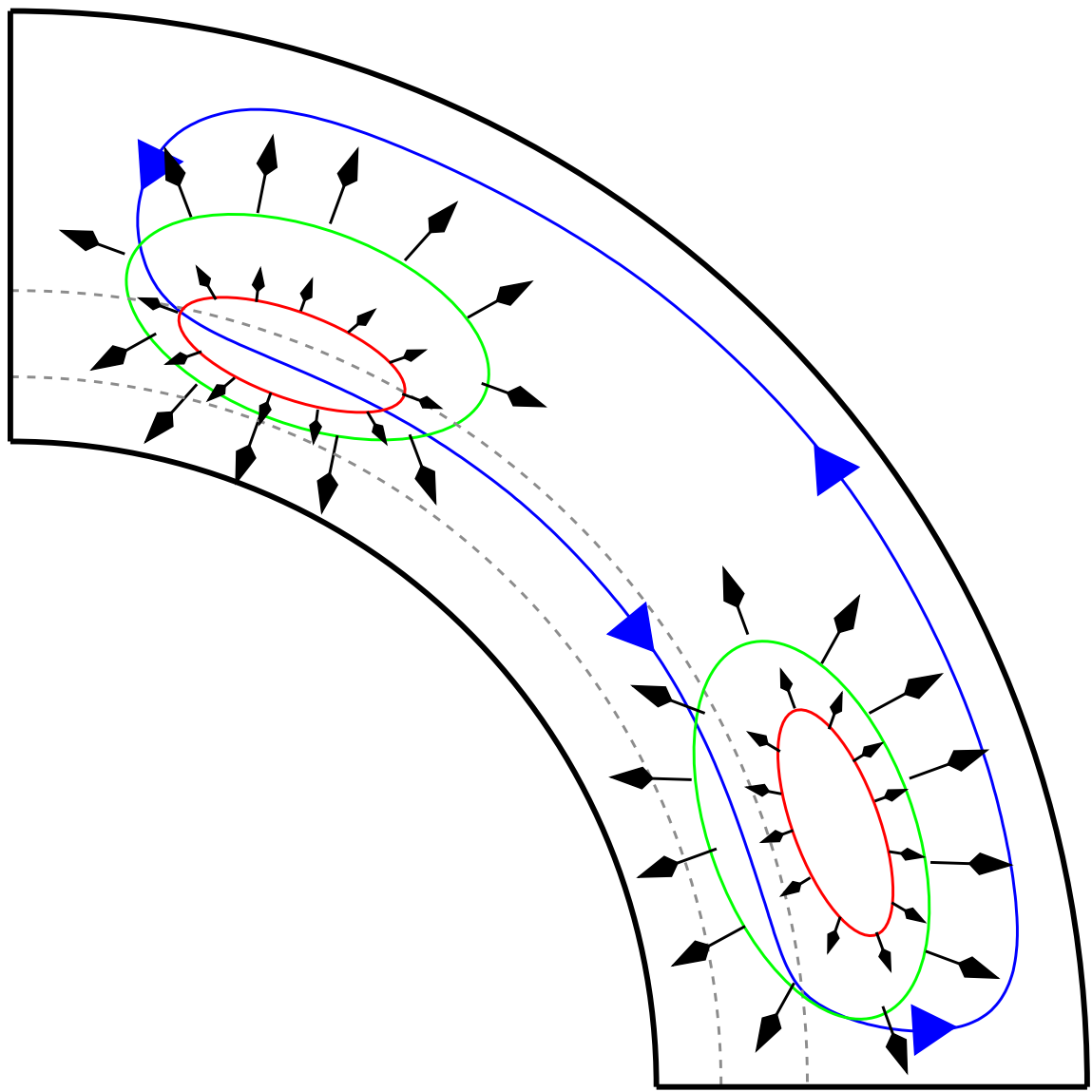}
\caption{
Schematic diagram shows how, with the help of downward advective 
transport, the high-latitude toroidal fields (cross section of which
is represented by the red loop, direction being perpendicular to the 
plane of the paper) can access the shear layer (within gray 
doted lines) is more efficiently amplified when there is strong 
$\eta$-quenching. This can be compared to the toroidal fields 
without $\eta$-quenching (cross-section is represented by the green 
loop, direction being perpendicular to that loop), which diffuse more
efficiently. By contrast, the upwelling flow at low-latitudes
makes the toroidal fields stay away from the shear layer when
diffusivity is more suppressed (see the low-latitude red-loop).
}
\label{schem}
\end{figure}

\clearpage

The contrasting feature of the field amplification at high and low
latitudes in the tachocline (see the dotted line and solid line in
the upper panel of Figure 7) occurs, respectively, due to the difference
in advective transport at high and low latitudes. If it would have
been due to the suppression of the diffusivity and hence a prolonged
shearing effect, we would have expected the increase in field amplitude
in all latitudes. Clearly the directional transport plays some role
here. We can understand this by using the schematic diagram shown in
Figure 8. The advective transport being downward at high-latitudes,
the poloidal fields there can reach the shear layer, for both cases
when there is no $\eta$-quenching and when
there is strong $\eta$-quenching. But in the latter case, 
the toroidal fields, originated from the sheared poloidal fields there, 
undergo less diffusion and stay at
the tachocline shear layer for longer time than in the former case.
The high-latitude toroidal fields (which are normal to the red loop in the
figure) undergo further amplification due to local feeding by 
new toroidal field lines that are created by shear due to the
strong tachocline differential rotation, compared  to the case without
$\eta$-quenching where the toroidal field undergoes more
diffusion (see the high-latitude green loop which represents
a section normal to a bunch of toroidal field lines without
$\eta-$-quenching). As a consequence, with the
increase in the $\eta$-quenching, there is a systematic increase in
the high-latitude toroidal field amplitude.

On the other hand, the upwelling flow near the equator 
pushes the low-latitude poloidal fields upward, away 
from the shear layer, making it difficult to create
toroidal fields at such latitudes. However a question
arises here: the low-latitude poloidal fields are advected
upward always, no matter whether the $\eta$-quenching is 
present or not, but why do the low-latitude toroidal fields
decrease with the increased $\eta$-quenching 
(with decreasing $B_q$), instead of being independent of $B_q$?
Again we take the help of the schematic diagram in Figure 8 
to explain this feature. This happens due to the combination 
of decrease in diffusivity and upward transport. 
In spite of the fact that the meridional flow at low latitudes
always takes the poloidal fields away from the radial shear 
layer, there is a small amount of toroidal field being produced
there and  another amount produced at the convection zone due to
latitudinal shearing. Given that the only mechanism that is able to 
transport the toroidal field downwards, at low latitudes, 
is the diffusive transport, in the case without $\eta$-quenching 
(green loop), the toroidal field produced in the convection 
zone expands and reaches a portion of the tachocline. 
There, it encounters the existing amount of toroidal field and thus,
increases its magnitude. However, in the case with $\eta$-quenching,
the toroidal field produced in the convective zone remains confined 
to a small region (red loop) and the toroidal field at the tachocline
is  not effectively increased. Hence the decrease in field 
amplification at low latitudes happens with increased $\eta$-quenching.

\subsection{Influence of $\eta$-quenching in butterfly diagram
and cycle period}

The increase in high-latitude toroidal fields and decrease in low
latitude toroidal fields in the tachocline regions, with smaller
$B_q$, will influence the butterfly  diagram accordingly. We recall
again that the contours of the toroidal field that appear in the
diagrams are computed from the radial average over the overshoot
region.

Figure 9 shows several butterfly diagrams for different values of $B_q$.
It can be seen how the butterfly wings, that are predominantly concentrated
within latitudes $\le 45 \degree$  when $B_q$ is large (less quenching),
move to higher and higher latitudes as $B_q$ decreases (more quenching).
This happens for the same reason that low-latitude fields at the tachocline
decrease whereas the high-latitude fields increase with enhanced
$\eta$-quenching. If a very large suppression in the $\eta$ occurs,
the butterfly diagram produced from this model will not be in accordance
with observations. The obvious question arises whether we can estimate 
how much $\eta$-quenching should be expected. \citet{brandenb01} showed in a
direct numerical simulation of dynamos that the saturation level of the
magnetic field is often close to the equipartition field strength
given by, $B_eq= \sqrt{\mu_0\, \rho\, u^2}$ in MKS unit, or $=\sqrt{ 1/2 \,
\rho\, u^2}$ in CGS unit, where $u$ is the rms value of the turbulent 
velocity, $\mu_0$ is the conversion factor between CGS and MKS unit 
and $\rho$ is the plasma density. With approximate values of the 
turbulent velocity of 5000 ${\rm cm}\,{\rm s^{-1}}$ at the base 
of the convection zone and density of 0.2 gm/cc, the equipartition
magnetic field comes out to be approximately $10^4$ Gauss. Assuming
that the back-reaction of the magnetic fields to quench the $\eta$ 
will not start until the equpartition field strength is reached,
the butterfly diagram will not depart much from observations.  
By contrast, in the present calculations large departures from the
observed butterfly diagram occur for $B_q \lesssim 500 $Gauss (see
Fig. 9d), but according to the above arguments, $\eta$-quenching
should not occur for fields so far below equipartition.

Note also that in the bottom panel of Fig. 9 the
toroidal field shows some small-scale structures, as we saw in the
toroidal field patterns in Figure 6. We repeat here that the primary
reason for the formation of these fragmented, small-scale structures 
is the competition between the two transport effects; the fields tend
to remain more frozen due to the lowering in the diffusivity while
the equatorward advective drag is pulling them, eventually causing 
their fragmentation. In order to check whether these are merely the
numerical effects due to resolution problem, we have performed three
experiments, namely doubling the resolution (experiment \#1), 
reducing the advective speed to half of the value used in the present 
paper (experiment \#2) and doubling the advective speed (experiment \#3). 
The results (figures not included) indicate that the fragmented, 
small-scale structures do not go away with the increased resolution; 
nor do they go away when the meridional flow-speed is reduced, but 
those structures are almost gone when the flow-speed is doubled, because
advection wins the competition in that case. Perhaps 
a minor contribution into those small-scale structures comes from the 
numerical effect due to model diffusivity reaching the limit of 
grid-diffusion, but our experiment \#3 confirms our  
physical explanation that the formation of those small-scale structures 
are the consequences of the two aforementioned competing processes. 

With smaller values of $B_q$, it is
possible to reach larger magnetic fields; however the system does
not reach a steady state.

The changes in the butterfly patterns indicate that these diagrams
remain consistent with observations up
to a certain increase in the $\eta$-quenching, but the influence of
an enhanced $\eta$-quenching is to make the butterfly-diagrams depart
further from the observations. The shift in the butterfly wings
towards higher latitudes could cause another problem in the dynamo
model, namely loss in the coupling between the two hemispheres across the
equator and hence, a shift to the quadrupole parity in the solution
when solved in a full spherical shell. In that case, to restore the
observed dipolar parity of the large-scale solar magnetic fields,
the help of additional downward transport using turbulent pumping
may be required (Guerrero \& de Gouveia Dal Pino 2008). We leave
those studies for the future.

In Fig. 9 we also see an increase in the dynamo cycle period with
increasing $\eta$-quenching, i.e. with decreasing $B_q$. The number of
butterfly wings produced in $60$ yr decreases. Figure 10 shows a plot
of dynamo cycle period as function of $B_q$, and the decrease in
cycle period with increased $\eta$-quenching is very clear.
This happens due to the competition between the flux-freezing effect
by the smaller diffusion and advective drag due to meridional flow.
The suppression in $\eta$ due to quenching works against the advective
transport of flux.

\clearpage

\begin{figure}[hbt]
\epsscale{0.7}
\plotone{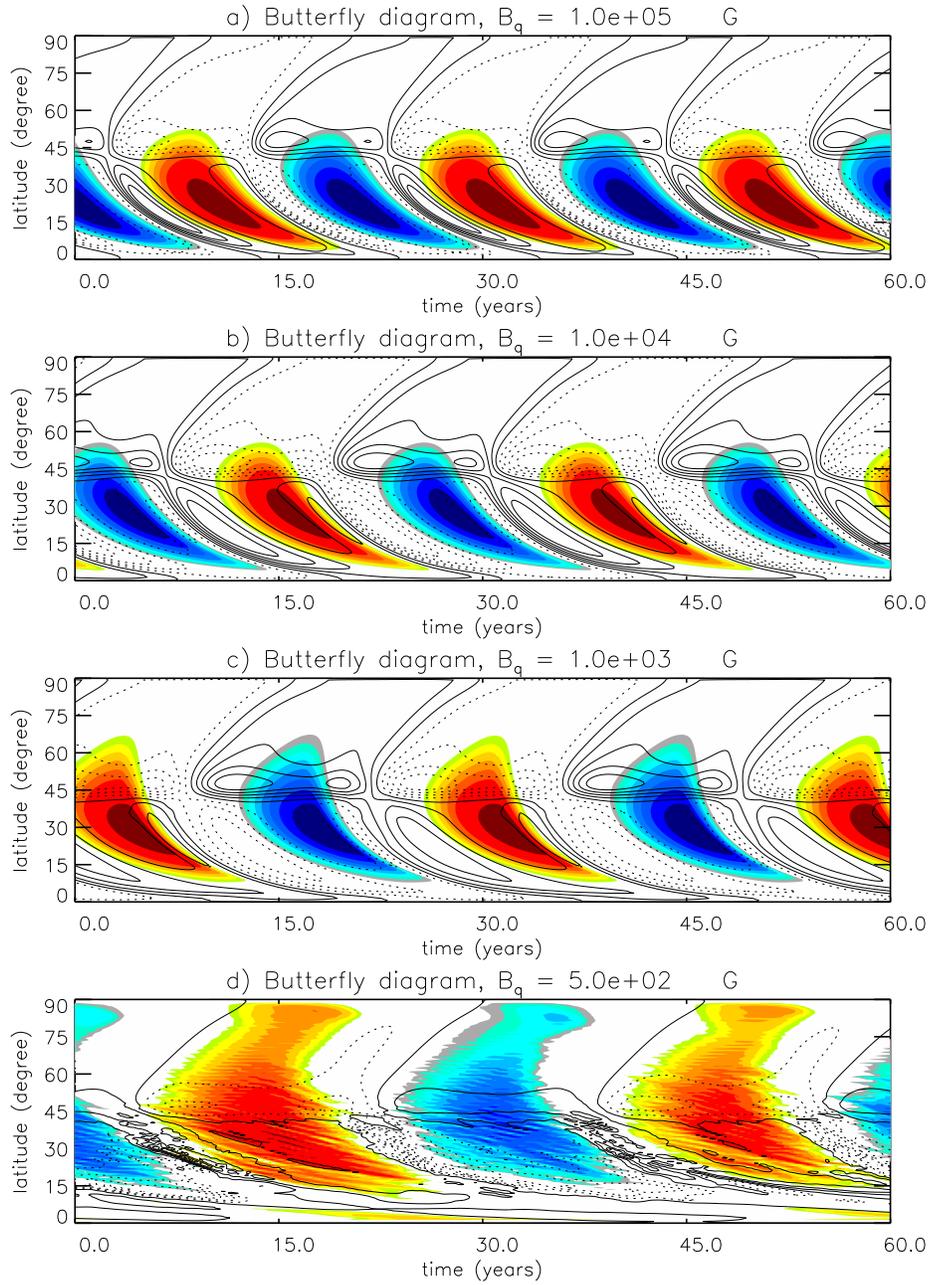}
\caption{
Butterfly diagrams for different values of $B_q$. Top:
  $B_q$$=$$10^5$ G; second from top: $B_q$$=$$10^4$ G; third from top:
  $B_q$$=$$10^3$ and bottom:  $B_q$$=$$5 \times 10^2$ G.
}
\label{fig6}
\end{figure}

\clearpage

\subsection{Advection-dominated versus diffusion-dominated dynamos}

In Babcock-Leighton dynamo models, the meridional flow is the conveyor
belt that carries the magnetic field, both at the surface in order to
produce a new poloidal field, and at the bottom of the convection zone
where it transports the toroidal fields in the direction of the
equator in such a way that this flow dominates over other parameters
in setting the period of the cycle \citep{dc99}. But for this
process to occur the advective term in eq. \ref{eq1} must
dominate the diffusive term in determining the time-scale of the
system. Thus this class of models require values of
$\eta_{cz}$$\lesssim$$2\times10^{11}$ cm$^2$ s$^{-1}$.
However, the values inferred from the mixing length theory at the
surface are one to two orders of magnitude larger than the values
considered in the bulk of the convection zone in the advection-dominated
flux-transport dynamo models. To our knowledge, until now
there is no accurate estimation of the magnetic diffusivity at
the convection zone and beneath. This limitation have recently led to
criticisms to the flux-transport scenario (see more about this
discussion in \citet{char07,yetal03}).

In previous sections, we have described the action of the
$\eta$-quenching in a Babcock-Leighton dynamo and found that, depending
on the quenching parameter $B_q$, the diffusivity can be locally
suppressed by up to three orders of magnitude and this effect can also
increase the cycle period (Figure 10). This is an indication that the
average radial value of $\eta_{cz}$ is also being suppressed. One question
arises here: will diffusion-dominated dynamos, which in general
produce much faster cycles, change to advection-dominated
dynamos due to the suppression of $\eta$ by quenching mechanism,
and produce a cycle period similar to the observed sunspot cycle?

\begin{figure}[hbt]
\epsscale{1.0}
\plotone{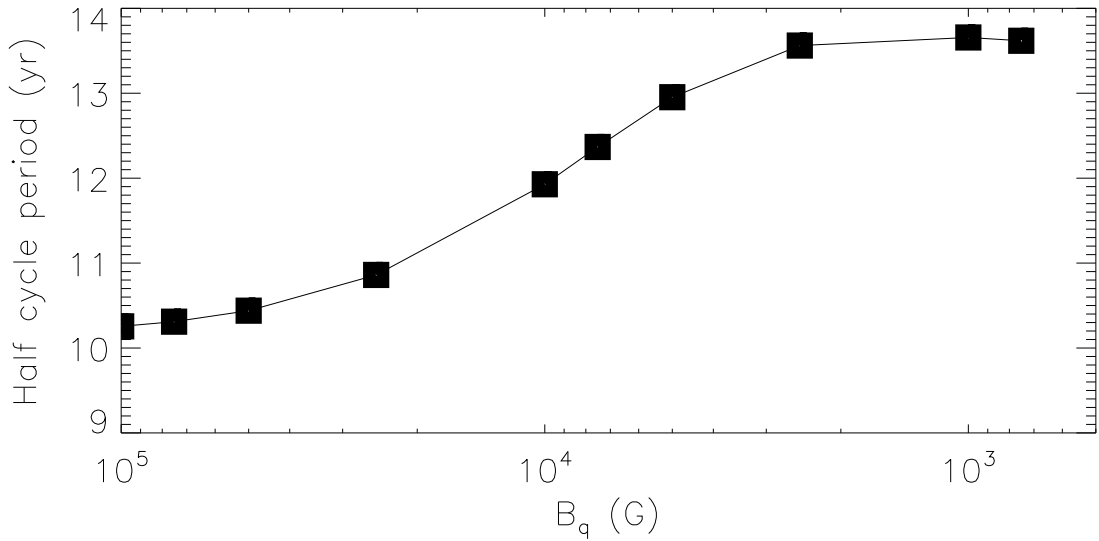}
\caption{
Half-cycle period as function of $B_q$.
}
\label{fig7}
\end{figure}

\begin{figure}[hbt]
\epsscale{1.0}
\plotone{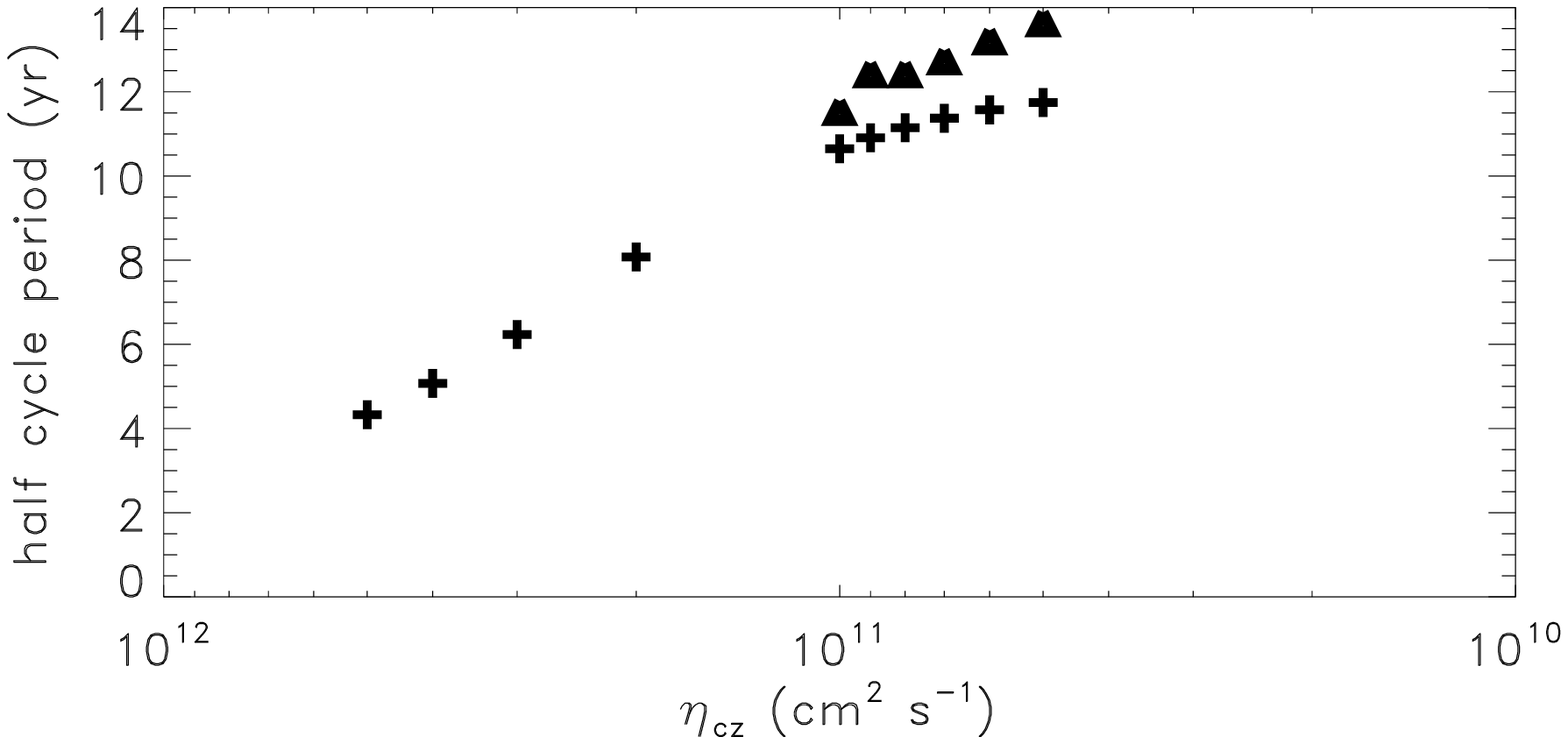}
\caption{
Half-cycle period as function of $\eta_{cz}$ for
  $B_q$$=$$10^3$ G (triangle symbol) and $B_q$$=$$10^4$ G (cross symbol).
}
\end{figure}

We have performed simulations to try to answer this question. We have 
increased progressively the
diffusivity from $5\times 10^{10}$ cm$^2$ s$^{-1}$ (the value
employed in the previous calculations) to the largest allowed value,
namely $5\times 10^{11}$ cm$^2$ s$^{-1}$, for which we obtain an
oscillatory solution (not decaying due to large diffusivity). The
latter case is the diffusion-dominated regime, and the cycle period
is small, determined by the diffusivity values. We then performed
the same numerical experiments as before by switching on the
$\eta$-quenching, for the two representative values of $B_q$, $10^3$
and $10^4$ G and maintained the constant dynamo efficiency defined
by  $C_{\Omega}C_{\alpha}$, for all the simulations. We note that
keeping $C_{\Omega}C_{\alpha}=\alpha_0 \Omega_{eq}
R^3/\eta_{cz}^2=cte$ implies a change in $\alpha_0$ as
$\eta_{cz}$ changes (see the Table 1 for the values used for
$\alpha_0$ for each simulation, as well as  the maximum values for
the average toroidal field at the base of the convection zone,
$\overline{B_{r_c}}$, and the radial field, $B_r$, at the surface).

In Figure 11, we present the dynamo cycle period as function of
magnetic diffusivity for the two cases, with $B_q=10^4$ G (pluses)
and $B_q=10^3$ G (triangles). The models with $B_q$$=$$10^4$ G
(plus symbols) present two different regimes for the slope of the
curve of the period versus $\eta_{cz}$. For values of $\eta_{cz}$$\le$
$10^{11}$ cm$^2$ s$^{-1}$, the cycle period does not vary much
with the value of $\eta$; it is primarily determined by the meridional
flow speed. For $\eta_{cz}$$>$$10^{11}$ cm$^2$ s$^{-1}$, the cycle
period varies more rapidly with $\eta_{cz}$, indicating that the
models are operating in the the diffusion-dominated regime.

\begin{figure}[hbt]
\epsscale{1.0}
\plotone{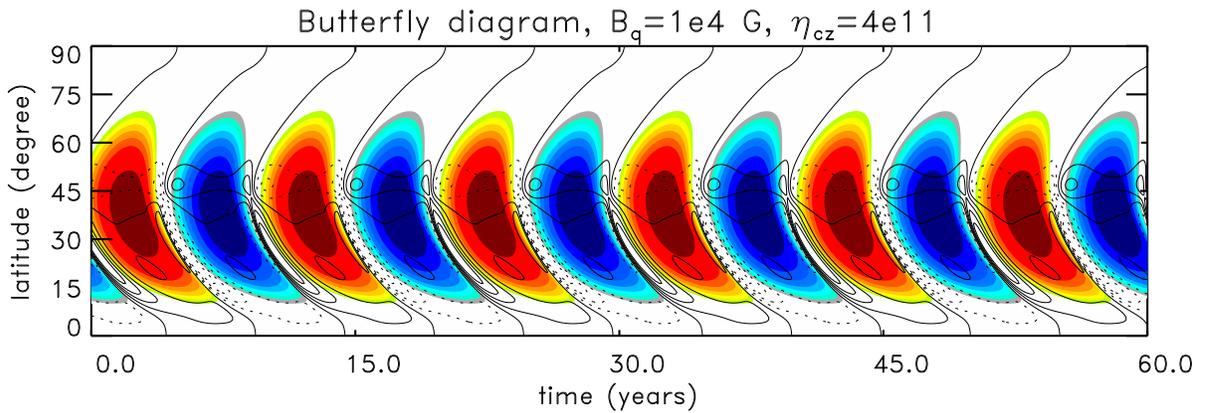}
\caption{
Butterfly diagram for the model with $B_q$$=$$10^4$ G and
  $\eta_{cz}$$=$$4 \times 10^{11}$ cm$^2$ s$^{-1}$.
}
\label{fig11}
\end{figure}

For the case of $B_q$$=$$10^3$ G, the highest value of $\eta_{cz}$,
that allows a steady state solution with a well defined period is
$10^{11}$ cm$^2$ s$^{-1}$. In this case, the diffusivity is highly
intermittent with larger gradients than in all the previous
calculations. However, we find from the plot of triangles in Figure 10,
a smooth change of the period as function of $\eta_{cz}$.

Figure 12
shows a butterfly diagram for one of these diffusion-dominated cases.
For these models (with higher $\eta_{cz}$), we also see in Figure 12
that the phase difference between the toroidal field at the lower
latitudes and the radial fields near the poles is $\sim \pi$ rather
than $\pi/2$.

These results clearly reveal that a diffusion-dominated dynamo with
meridional circulation remains in a diffusion dominated regime even
if the diffusivity is locally suppressed by the back-reaction of
magnetic fields. However, we emphasize the fact that the amplitude
and profile of the magnetic diffusivity inside the convection zone
are still unknown.

\clearpage

\begin{table}
\begin{center}
\caption{Simulation parameters ($B_q$, $\eta_{cz}$ and $\alpha_0$)
and model-output (maximum toroidal field at convection zone base,
maximum surface radial field and dynamo cycle period).\label{tbl-1}}
\begin{tabular}{crrrrrrrrrrr}
&&&&& \\
&&&&& \\
\tableline\tableline
$B_q$ (G) & $\eta_{cz}$ (cm$^2$ s$^{-1}$) & $\alpha_0$ (cm s$^{-1}$)  &
$B_{max}$ (G) & $B_{r_{max}}$ (G) & $T/2$ (yr)\\
\tableline
$10^3$ & $5\times10^{10}$ & $50$ & $4.8\times 10^4$ & $153.7$ &
$13.65$ \\
$10^3$ & $6\times10^{10}$ & $72$ & $5.0\times 10^4$ & $170.0$ &
$13.23$ \\
$10^3$ & $7\times10^{10}$ & $98$ & $7.5\times 10^4$ & $543.1$ &
$12.75$ \\
$10^3$ & $8\times10^{10}$ & $128$ & $7.9\times 10^4$ & $832.8$ &
$12.44$ \\
$10^3$ & $1\times10^{11}$ & $200$ & $9.3 \times 10^4$ & $981.428$ &
$11.55$ \\
\tableline
\tableline
$10^4$ & $5\times10^{10}$ & $50$ & $3.6\times10^4$ & $121.9$ &
$11.74$ \\
$10^4$ & $6\times10^{10}$ & $72$ & $4.5\times10^4$ & $174.1$ &
$11.57$ \\
$10^4$ & $7\times10^{10}$ & $98$ & $5.3\times10^4$ & $252.2$ &
$11.37$ \\
$10^4$ & $8\times10^{10}$ & $128$ & $6.1\times10^4$ & $345.2$ &
$11.14$ \\
$10^4$ & $9\times10^{10}$ & $162$ & $6.7\times10^4$ & $452.0$ &
$10.90$ \\
$10^4$ & $1\times10^{11}$ & $200$ & $7.2\times10^4$ & $572.4$ &
$10.64$ \\
$10^4$ & $2\times10^{11}$ & $800$ & $9.5\times10^4$ & $2547.5$ &
$8.07$ \\
$10^4$ & $3\times10^{11}$ & $1800$ & $1.0\times10^5$ & $5998.7$ &
$6.23$ \\
$10^4$ & $4\times10^{11}$ & $3200$ & $1.1\times10^5$ & $10635.7$ &
$5.07$ \\
$10^4$ & $5\times10^{11}$ & $5000$ & $1.2\times10^5$ & $16515.5$ &
$4.32$ \\
\tableline
\end{tabular}
\end{center}
\end{table}

\clearpage

\section{SUMMARY AND COMMENTS}

We have explored here the effects of diffusivity quenching on
Babcock-Leighton flux-transport solar dynamo models. We used as
initial condition a converged solution of a dynamo model that
reproduces most of the main features of an observed solar butterfly
diagram. The $\eta$-quenching  was then included in the model
through an algebraic function that is similar to the usual
$\alpha$-quenching formula. After some years of evolution, the
system reaches a new steady state configuration in which the
poloidal and toroidal components of the magnetic field, $A$ and $B$,
respectively,  as well as the magnetic diffusivity, $\eta$, exhibit
a cyclic behavior.

With the new $\eta$ profile (Equation 5), the decrease in the
diffusivity can be as large as three orders of magnitude, but it is
not homogeneous over the whole domain since it presents a pattern with
peaks and valleys that anti-correlate with the spatial
distribution of the amplitude of the toroidal fields - the smaller
the $B_q$ (the value of the magnetic field at which the diffusivity
begins to be quenched; eq. 5) the narrower the final $\eta$ profile
(Figs. \ref{fig8} and \ref{fig9}). This spatial fluctuation in the
magnetic diffusion results in the formation of small and long-lived
regions of concentrated magnetic field that appear predominantly at
the equatorial part of the overshoot tachocline as well as in the
middle of the convection zone. The role of the meridional flow is
very important in this result, since the toroidal fields have time
enough to increase and be dragged along with the poloidal
field lines before being dissipated. Note that if these strong flux
tubes produced at the middle convection zone emerge to the surface
they may not agree with the Joy's law.

We have found that, contrary to the effects of $\alpha$-quenching, 
which shows an almost linear coupling between the saturation field, 
$B_0$ and the final value of $\alpha$ (see Fig. \ref{fig4}), the dependence
between the diffusivity saturation field, $B_q$, and the final value
of $\eta$ is predominantly non-linear, especially at the base of the
convection zone where both the radial and the latitudinal shear are
competing with the advective transport and with diffusive spreading. It
was found that at the tachocline ($r$$=$$0.7\sr$) the magnetic field
can be amplified by a factor up to $\sim2.5$ at the highest
latitudes, while for the equatorial regions the magnetic field
decreases by approximately the same factor. On the other hand, at
the center of the convection zone ($r$$=$$0.8\sr$), the magnetic
field for both low and mid-latitudes have the same increase, by
a factor as large as $\sim 2$ with respect to the no-quenching case.

The consequence of these effects on the butterfly diagram is the
increase (decrease) of the toroidal fields at the high (low)
latitudes and the increase of the cycle period for an enhanced
$\eta$-quenching. This new distribution of magnetic fields with
latitude not only shows a gradual departure from observations, 
but also could cause other problems in the dynamo models, such as the
loss of coupling between hemispheres when solved in a full spherical
shell, resulting in a quadrupole parity solution.

Since the period of the cycle diminishes when the efficiency of the
quenching increases, we have also explored if the $\eta$-quenching
can make a diffusion dominated dynamo, which is characterized by a
small cycle-period, to evolve to an advection dominated dynamo. We
have found that a diffusion-dominated dynamo with a meridional flow
remains in the diffusion regime even when the diffusivity is locally
suppressed due to the strong magnetic fields.

The results summarized above indicate that in the scenario of a pure
Babcock-Leighton dynamo, with a meridional flow operating as a
conveyor-belt and strong magnetic flux tubes emerging from the base
of the convection zone, the turbulent diffusivity is probably weakly
suppressed, i.e., it is quenched only for high values of the
magnetic field. This implies that its role in the amplification of the
magnetic field to values above the equipartition field should not be
significant enough. Notice, however, that other effects, like
turbulent pumping, which have been demonstrated to be important in
the dynamo operation \citep{gdp08}, have not been considered here.
The contribution of turbulent pumping might, for instance, result in a
different transport of the magnetic fields, changing the parts of
the parameter space where
the $\eta$-quenching would become dominant. This will be explored in
forthcoming work. Finally, it is also important to remark that the
quenching effects of the diffusivity still need to be explored  in
other classes of dynamo models, for example, the ones operating
with a surface shear layer. These will also be considered in future
work.

\acknowledgments We thank E. J. Zita, Matthias Rempel and Peter
Gilman for helpful discussion on this work. We extend our thanks to
an anonymous referee for a very thorough review and for many 
constructive comments and criticism on an earlier version of this
paper -- incorporating them in the revised manuscript has significantly
improved the paper. G. G and E.M.G.D.P 
acknowledge partial support from grants of the Brazilian Science 
Foundations FAPESP and CNPq. This work is partially supported by
NASA grant NNX08AQ34G. National Center for Atmospheric Research
is sponsored by the National Science Foundation.

\end{document}